**Creation of inclusive spaces with astromimicry**
Jorge Moreno

***The universe can inspire us to design communities that foster equity and inclusion.***

Coined in 1997 by Janine Beynus, the word 'biomimicry' describes a method in which nature inspires us to address challenges in design and innovation. Notably, Leonardo Da Vinci's conceptualisation of vehicles that can fly like birds comes to mind. Other examples include ventilation systems inspired by ant nests and wind turbines influenced by whales. In this essay I propose astromimicry: a framework where humans can seek inspiration from the Universe to design more inclusive organisations.

To illustrate, consider the use of binary language in the field of galaxy evolution. Galaxies are classified as either spiral or elliptical, red or blue, passive or star-forming, and so forth. Although appealing, this framework fails to fully describe galaxies that do not conform to these simplistic demarcations. Inspired by this example, we can acknowledge the harm caused by societal binaries, which have been used as instruments of violence by demagogues throughout history. With astromimicry, we are inspired to have more nuanced conversations about our identities. We are encouraged to challenge the discrimination of 'the other'. Along this vein, the separation of galaxies into centrals and satellites also invites us to reflect on the harm inflicted by borders and citizenship status on many of our colleagues.

In the classroom, the students and I have also enquired the Universe for ideas pertaining to equity and inclusion. One example is the question of how to form groups. If students choose to work with the people they know, segregation might occur. However, if groups are assigned randomly, minoritised students might face discrimination and exclusion. To solve this, we were inspired by the Magellanic clouds, two tiny galaxies orbiting the Milky Way. By being in a pair, these galaxies can support each other whilst simultaneously benefiting from their membership to the Milky Way group. Correspondingly, groups are formed randomly but students are allowed to follow a classmate (with mutual consent) if they need extra support.

Another example comes from molecular clouds. The formation of interstellar molecules is aided by dust. Dust particles can foment molecular formation on their surfaces and shield these fragile molecules from external ultraviolet radiation. Using astromimicry, we recognise that students with privilege can act as dust particles, by humbly supporting their marginalised classmates whilst protecting them from external challenges.

Finally, consider the story of the two recently unveiled galaxies lacking dark matter, whose discovery inevitably posed a significant challenge to our understanding of galaxy evolution. Using a novel computer simulation, my collaborators and I recently demonstrated that these peculiar galaxies can exist naturally in a Universe that follows the standard cold dark matter paradigm (Moreno et al. 2022, Nature Astronomy, https://www.nature.com/articles/s41550-021-01598-4). We identified seven such galaxies in our simulation, which we named in honour of the seven Cherokee clans. As a person of Indigenous ancestry, I hold these galaxies close to my heart. In general, the interaction between a small and a giant galaxy can have one of two outcomes: either the satellite assimilates into the dominant neighbour via merging, or it is completely destroyed by the gravitational field of the host. These seven galaxies follow a third path: they bravely confront their Goliath multiple times. And although they pay a heavy price by losing most of their dark matter, their stellar core remains.

This tale is the story of every person of Indigenous ancestry in the world. For the Cherokee Nation, they are still suffering from the aftermath of deportation under the Indian Removal Act, and their subsequent internal struggle to decide whether to fight, hide or assimilate. As for me, it is very can be confusing and painful to carry a European name and gender, and to only speak European languages. It is heartbreaking not to know the language, music, cuisine, customs, or cosmogony of my ancestors. Nevertheless, I am still here, and I am thriving. My stellar core remains, and I am living my ancestors' wildest dreams.

With astromimicry, we can be inspired to design communities where vulnerable members are not annihilated nor forced to assimilate. Imagine a culture where folks are not expected to be "assertive", to behave in ways frequently associated with masculinity or any other gender assigned to them, or to perform neurotypical behaviour. A place where they do not need to suppress their dialect, or accent, and adopt a "white voice" to be treated seriously. A space where they can wear their natural hair, a hijab, their traditional attire, clothes that align with their gender (if they have one), or any style they desire. An organisation where the various forms of discrimination — based on gender, race, sexuality, class, disability, body shape, age, citizenship status, police record, veteran status, housing situation, trauma, mental illness, or chronic pain — are addressed and its members are protected and given the resources and support they need to heal. Imagine a community where its members are not asked to shed their layers and dehumanise themselves to survive. As we re-imagine and re-invent the ways we organise ourselves, let us look up, not down.

Jorge Moreno is in the Department of Physics and Astronomy, Pomona College, Claremont, CA 91711, USA

Correspondence. jorge.moreno@pomona.edu.
Competing Interests. The author declares no competing interests.

**Acknowledgements**
The author thanks Nicole Arulanantham, James Bullock, Kate Daniel, Shany Danieli and Camila Moreno for their comments on an earlier draft – and Jessica Kizer, Katrina Miller, Camila Moreno, Mateo Moreno, Viridiana Moreno, Rosalia Romero and Asha Srikantiah for their wisdom and the conversations that helped shape this essay. The author is also grateful to Asha Srikantiah for suggesting the word "astromimicry" to describe the framework proposed here. This work was conducted on Tongva-Gabrielino land.